\documentclass[runningheads]{llncs}
\usepackage[T1]{fontenc}
\usepackage{url}
\usepackage{hyperref}
\usepackage{graphicx}
\begin{document}
\title{Relational AI in Education: Reciprocity, Participatory Design, and Indigenous Worldviews}

\author{Roberto Martinez-Maldonado\inst{1} \and
Vanessa Echeverria\inst{2} \and
Jenna Hawes\inst{3}\and
YJ Kim\inst{4} \and
Zara Maddigan\inst{3} \and
Mikaela Milesi\inst{1} \and
Todd Nelson\inst{3} \and
Yi-Shan Tsai\inst{1}
}
\authorrunning{Martinez-Maldonado et al.}

%
\institute{Monash University, Australia 
\email{[roberto.martinezmaldonado,mikaela.milesi,yishan.tsai]@monash.edu} \and
RMIT, Australia  
\email{vanessa.echeverria@rmit.edu.au}\and
Ninti One, Australia
\email{[jenna.hawes,zara.maddigan,todd.nelson]@nintione.com.au} \and The University of Adelaide, Australia
\email{yj.kim@adelaide.edu.au}\\}

\titlerunning{Relational AI in Education}
%

%
%

%
\maketitle              
\begin{abstract}

Education is not merely the transmission of information or the optimisation of individual performance; it is a fundamentally social, constructive, and relational practice. However, recent advances in generative artificial intelligence (GenAI) increasingly emphasise efficiency, automation, and individualised assistance, risking the weakening of relational learning processes. Despite growing adoption, AI in education (AIED) research has yet to fully articulate how AI can be designed in ways that sustain the social and ecological relationships through which learning occurs. In this paper, we re-centre education as \textit{relational} and frame learner–AI interactions as context-specific relationships with clearly defined purposes and boundaries, rather than positioning them as substitutes for, or replacements of, human interaction. Grounded in participatory design practices and inspired by Indigenous worldviews (including Aboriginal Australian, Native American, and Mesoamerican traditions) that foreground reciprocity and relational accountability, we argue that meaningful educational AI should support learning \textit{with} others rather than replace them. We advance this perspective by: i) conceptualising AIED as a relational design problem grounded in reciprocity; ii) articulating key tensions introduced by GenAI in education; and iii) outlining design directions that expand the AIED design space toward reciprocity, including when not to use AI, how to define pedagogical boundaries, and how to support responsible uses of AIED innovations that sustain communities and natural environments.


\end{abstract}
\section{Introduction and Background}
Education has long been understood as a social, relational, and constructive practice, rather than a process of information transmission or individual performance optimisation \cite{freire1970pedagogy,biesta2015beyond}. Accordingly, learning emerges through interactions among learners, educators, communities, artefacts, tools, and environments, and is shaped by dialogue, negotiation, and shared understanding. From this perspective, knowledge is not simply acquired but co-constructed, situated within cultural practices, histories, and interactions with others \cite{engestrom1987learning,NRC2000LearnersLearning}.

Recent advances in artificial intelligence (AI), including generative AI (GenAI), challenge relational perspectives of education by emphasising efficiency, automation, and individualised assistance \cite{Chugh2025,Selwyn2019Robots}. While such systems can support learning and professional performance in specific contexts, their uncritical use risks weakening fundamental learning processes, including critical reflection and productive struggle \cite{Fan2025beware,Lee2025cognitive}. Productive struggle refers to the sustained effort invested in challenging tasks that builds deep understanding and reasoning skills rather than shortcuts to correct solutions \cite{young2024productive}, a critical opportunity for learning that may be diluted when AI systems minimise or bypass such effort \cite{alley2025struggle}.

In addition, when learning is positioned primarily as interaction between an individual and an AI system, the social fabric of education, such as peer collaboration, mentorship, and collective sense-making, can become secondary or diminished unless systems are intentionally designed to promote peer learning \cite{Sichterman2025}. Emerging empirical evidence underscores this tension: although interaction with GenAI can enhance immediate task performance, these gains may not persist when learners subsequently work independently and may be accompanied by reduced motivation and increased disengagement when AI use is framed primarily as individual work \cite{Wu2025}.

In response to these tensions, re-centring education as relational invites the design of learner–AI interactions as context-specific, with clearly defined purposes and boundaries, foregrounding the situated social, educational, professional, and cultural contexts in which AI is used. While AI in education (AIED) research rarely frames AI as a all-purpose replacement for human interaction \cite{Sharples03072023}, broader public discourse and everyday use can position AI as a convenient alternative for tasks such as explanation, feedback, and problem solving. This reveals a tension between research intentions and emergent practices. Our concern, therefore, is not with a singular stance in the field, but with how default uses of AI may gradually reconfigure learning as an individualised interaction with technology rather than a relational practice involving others. Although AI can effectively support individual learners, for instance through rapid access to information and personalised recommendations, overreliance on such interactions risks diminishing the role of peer collaboration and shared meaning-making, which are central to learning \cite{Stahl2006GroupCognition}.

A relational design orientation can respond to these challenges by embedding AI within collaborative learning arrangements; for example, deploying AI agents to prompt reflections, present alternative perspectives, or mediate learner–learner interaction, rather than positioning AI as a private, individual interlocutor \cite{jin2025machines}. Moreover, designing learner–AI interactions requires taking a clear stance on how knowledge should be accessed, when it should become available, and under what conditions it should be produced. Many contemporary technologies, such as the internet and GenAI chatbots, are built on an assumption of instant, on-demand access: information is expected to be searchable, retrievable, and generated immediately, often with minimal context, participation, or accountability. This design assumption normalises the idea that knowledge is universally available and frictionless. In contrast, some pedagogical traditions view understanding as something that develops progressively through guided participation, dialogue, and practice \cite{vygotsky1978mind,Toukan20102017}. Similarly, Indigenous perspectives emphasise that knowledge is relational and situated—shared within community, tied to responsibility, readiness, and context, and sometimes earned rather than instantly delivered \cite{Sinclair_2024,worrell2025indigenous}.

Extending this orientation further requires making visible the non-artificial dimensions of AI  (such as human labour, extractive data practices, energy consumption, and land-based resources), thereby situating learner–AI interactions within broader networks of social, material, and environmental responsibility \cite{Crawford2021Atlas}. Emerging Indigenous AI protocols articulate principles for such accountability, foregrounding relational, cultural, and material obligations that are often overshadowed by dominant design aims centred on efficiency, scalability, and optimisation \cite{Lewis2020IPAI}.

Yet, to meaningfully rethink AI through Indigenous worldviews, we need to move beyond a surface understanding of “\textit{relationality}” as simple connectivity. Graham’s \cite{graham2023law} concept of the “reflective self” reframes this: rather than the individual Cartesian ego (“\textit{I think therefore I am}”), the self is understood as communal and defined by location (“\textit{I am located therefore I am}”). This provides a stronger ethical and ontological grounding for our arguments in the paper about AI, obligation, and place-based accountability.

\begin{figure}[h]
    \centering
    \includegraphics[width=0.8\linewidth]{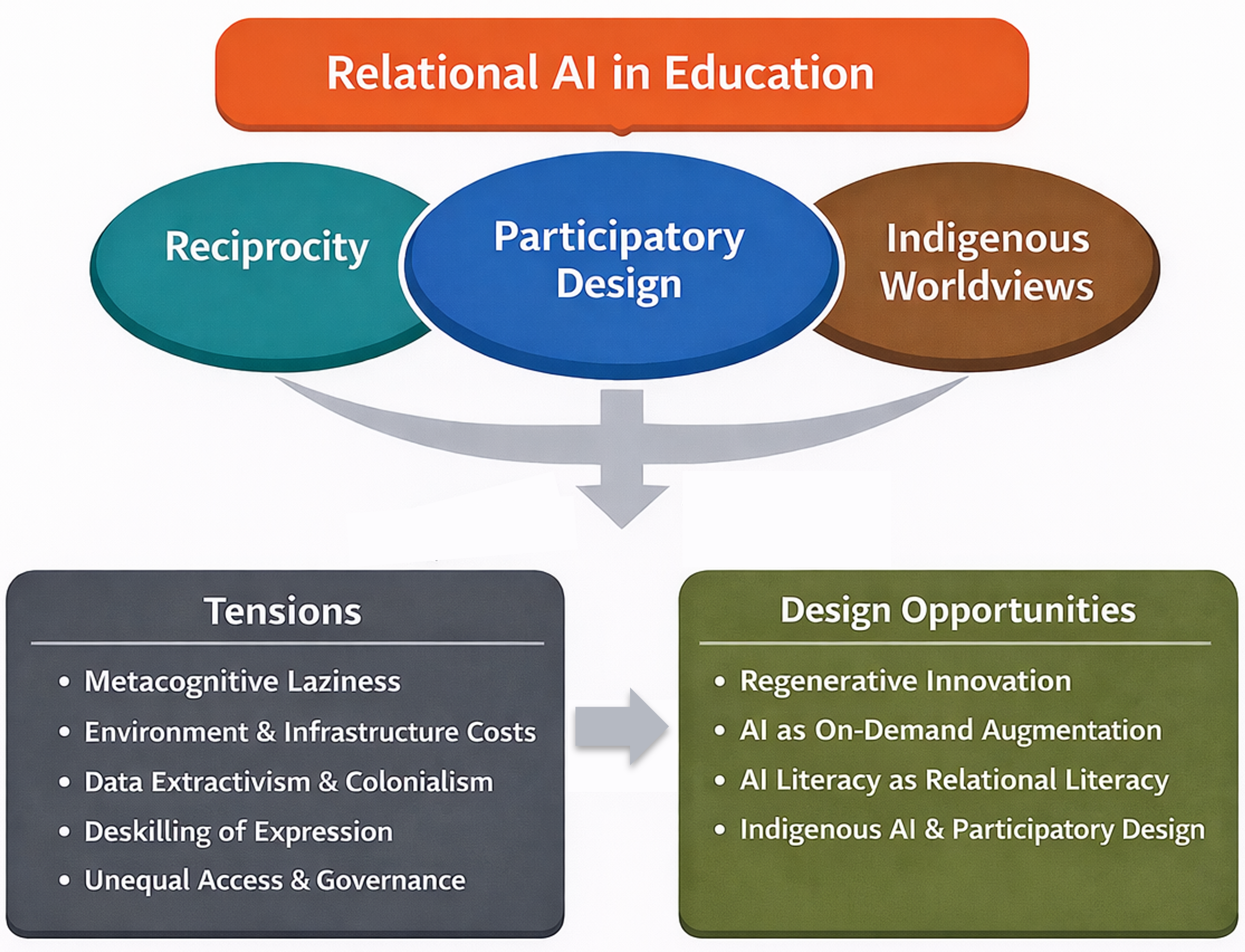}
    \caption{Conceptual framing of relational AI in education, linking reciprocity, participatory design, and indigenous worldviews to key tensions and design opportunities.}
    \label{fig:view}
\end{figure}

Grounded in participatory design practices and inspired by Indigenous worldviews (e.g., Aboriginal Australian \cite{Jones2025}, Native American \cite{Munroe04052022,Kimmerer2015Braiding}, and Mesoamerican Indigenous traditions) that foreground reciprocity and relational accountability, this paper argues that meaningful educational AI should support learning \textit{with} others, rather than replace human relationships, while strengthening reciprocal relationships with self, others, and the environment. We examine key tensions introduced by GenAI in educational contexts, including the displacement of social learning, the erosion of opportunities for reflection and productive struggle, and the obscuring of material and ecological costs. We then explore emerging opportunities for \textit{rethinking} the design of learner–AI interactions as reciprocal and outline potential future pedagogical directions that support responsible, context-sensitive, and community-sustaining uses of GenAI. Figure \ref{fig:view} provides an overview of the structure of this paper. 

Importantly, many of the tensions discussed below are not unique to GenAI, but reflect longer-standing tensions in technology-mediated learning. Concerns around individualisation and reduced social interaction have been present across AIED and educational technologies more broadly \cite{PorayskaPomsta2023EthicsAI}. From this perspective, GenAI amplifies existing dynamics rather than introducing entirely new ones. What remains insufficiently addressed, however, is how such systems can be designed to sustain reciprocity and participation as core conditions for learning.

\section{Positionality}
The authors come from diverse cultural and disciplinary backgrounds, including Ecuadorian, Mexican (Mesoamerican), Korean, Taiwanese, non-Indigenous Australian, and First Nations Australian perspectives. As researchers and practitioners working across community contexts, computer science, the learning sciences, learning analytics, AI in education, and human-centred design, our team brings together lived experience, professional practice, and technical expertise.


In this paper, we engage Indigenous worldviews and principles of reciprocity alongside human-centred design to problematise dominant perspectives on the use of AI in education and to reframe AI in education as a relational, community-embedded practice. We also recognise our varied positions within institutions in Australia—a nation that continues to wrestle with its colonial history and the ongoing, intentional marginalization and oppression of Aboriginal and Torres Strait Islander peoples and knowledges. This shapes both the possibilities and constraints of how we interpret, design, and advocate for AI in education.

\section{Contributions}
This paper makes three contributions:

\begin{itemize}

    \item We \textbf{conceptualise} AI in education as a relational design problem grounded in reciprocity.
    \item We identify \textbf{five tensions introduced by GenAI }(metacognition, environment, extractivism, expression, governance).
    \item We propose \textbf{four design directions }(regenerative innovation, on-demand AI, relational literacy, Indigenous participatory design).
\end{itemize}

\section{Tensions: Why Rethinking Is Needed}

\subsection{Metacognitive Laziness}
From a relational perspective, this tension concerns learners’ relationship with self, particularly their capacity for reflection and self-regulation. The term \textit{metacognitive laziness} has been introduced to describe a pattern in which learners increasingly offload planning, monitoring, evaluation, and reflective judgment to external systems, leading to diminished engagement in their own metacognitive regulation \cite{Fan2025beware}. In an experimental study comparing learners supported by GenAI, human experts, analytic tools, or no support, Fan et al. \cite{Fan2025beware} found that while GenAI improved short-term task performance, it was associated with altered self-regulated learning processes, including reduced evaluative and reflective engagement and greater dependence on AI. Performance gains did not translate into stronger knowledge transfer or sustained learning, suggesting that efficiency may come at the cost of deeper metacognitive engagement, which is central to learning \cite{Flavell1979Metacognition}. Developmental research further indicates that metacognitive skills are initially task-specific in young children and only begin to generalise reliably between approximately 8 and 13 years of age, continuing to develop through adolescence and plateauing in adulthood \cite{weil2013development}. Consequently, early and uncritical reliance on AI risks disrupting the development of these still emerging metacognitive capacities, potentially undermining learners’ ability to reflect on errors, regulate their learning, and develop independent learning skills over time.

\subsection{Environment and Infrastructure Costs}
Recent conversations among Aboriginal and Torres Strait Islander communities about AI futures emphasise that the use of AI cannot be separated from the health of \textit{Country} \cite{Barrowcliffe2025AICommunique}. 
In Indigenous Australian contexts, Country is a multifaceted and relational concept that extends far beyond land or territory: it can refer to the living connections between people, plants, animals, land, water, and sky, as well as to stories of creation, ancestral presence, seasonal cycles, and responsibilities passed across generations \cite{Murawin2021StateEnvironmentIndigenous}. From this perspective, AI is not understood as an abstract or immaterial technology, but as a set of infrastructures that materially affect Country through data centres, energy consumption, water use, and extractive (mineral) resource practices \cite{worrell2025indigenous}. This concern is particularly significant for Aboriginal and Torres Strait Islander peoples, whose knowledge systems understand learning, technology, and wellbeing as inseparable from relationships with land, kin, and future generations \cite{graham2023law}.

The rapid expansion of AI therefore intensifies long-standing tensions between technological development and environmental sustainability, raising urgent questions about who bears the ecological costs of digital innovation \cite{Crawford2021Atlas}. Yet, these impacts are not uniform: while large-scale model training and data centre infrastructure are particularly resource-intensive, inference costs vary widely depending on deployment context (e.g., cloud-based services versus edge or local models) \cite{DEVRIES20232191}. Moreover, AI should be understood within a broader ecosystem of digital technologies that can collectively contribute to environmental stress.

In education, this creates a profound challenge: how to benefit from AI-supported learning without contributing to forms of extraction that undermine the ecological and relational conditions that sustain learning itself. A relational and holistic educational approach foregrounds reciprocity as a guiding principle, recognising that relationships premised on extraction without return are ultimately unsustainable \cite{Kimmerer2015Braiding}. This suggests that AI in Education should be designed and used in ways that reciprocate rather than merely extract -- this is: supporting learning while sustaining the social, material, and environmental relations on which learning depends.


\subsection{Data Extractivism and Colonialism }
An extractivist mindset in AI operates not only through the intensive use of natural resources but also through data extractivism, in which data are treated as raw materials to be taken, aggregated, and repurposed without meaningful consent, governance, or reciprocity. Indigenous scholars argue that contemporary AI systems extend settler-colonial mindsets by extracting Indigenous data, knowledge, language, and cultural expressions—often invisibly and without authorisation—while presenting this process as neutral technological progress \cite{worrell2025indigenous}. From this perspective, data extraction mirrors historical resource extraction, reinforcing power asymmetries and undermining Indigenous data sovereignty: the right of Indigenous peoples to govern how data about their lives, communities, environments, and knowledge are collected, used, and shared. Extractivism in AI also depends on extensive forms of human labour, including data labelling, content moderation, and other invisible care work, frequently performed under precarious conditions and disproportionately borne by marginalised populations \cite{Crawford2021Atlas}. Environmental, data, and labour extraction are integral to how AI systems function and scale. 

While critiques of data extractivism often rely on economic terminology, Indigenous linguistics can more precisely capture the epistemic harm of these systems. Russ-Smith and Lazarus \cite{RussSmithLazarus2024} analyses the Wiradyuri word associated with artificial intelligence, \textit{bun-ngan} (made by another), and its links to \textit{bunambirra} (to sweep) and \textit{bunan} (dust). In this framing, AI is like a broom “\textit{sweeping all the dust or knowings out there in the world}”. This raises the critical question: Who is acknowledged in this dust? When AI sweeps up Indigenous and student knowledges as decontextualised “dust”, it severs them from their relational obligations and turns living wisdom into inert data.

In educational contexts, this raises critical challenges: how to design and justify AI-supported learning while remaining attentive to the extractive commitments embedded in AI systems themselves; and how to generate and share knowledge responsibly through the use of AI. From a Freirian perspective \cite{freire1970pedagogy}, when these commitments remain unexamined, AI risks reinforcing oppressive relations -- positioning learners as passive consumers of AI-generated content while obscuring the social relations through which knowledge is produced  -- rather than supporting education as a practice of freedom. Addressing this requires foregrounding the material, data, and labour dimensions of AI in educational discourse and cultivating relational awareness of the human, environmental, and power relations that sustain AI systems.

Extending this critique, an Aboriginal ethics perspective \cite{graham2023law} invites not only scrutiny of AI’s impacts but also interrogation of the developmental logic that drives technological systems. While recent work in AIED calls for frameworks that integrate fairness, accountability, transparency, and educational values into AI design \cite{Holmes2022,Bhimdiwala2022}, Aboriginal ethics shifts the focus from regulating outcomes to examining the orientation of innovation itself. As Graham \cite{graham2023law} argues, Western science and technology are propelled by an exploratory impulse and an accelerated trajectory toward goal-seeking that can gradually diminish the importance of ethical considerations regarding actions and intent. Within this logic, extractive practices can appear inevitable -- or even necessary -- for progress. In contrast, Aboriginal ethics foregrounds a custodial obligation: “\textit{looking after country}” and “\textit{looking after kin}”, embedding responsibility, relationality, and long-term balance into the very means by which outcomes are pursued. From this standpoint, the question is not simply how to mitigate AI’s harms, but how to reorient educational technology research and development toward custodial, relational, and intergenerational accountability.

\subsection{Long-Term Deskilling of Learner Expression}

Beyond metacognitive regulation, this tension also concerns learners’ relationship with self, particularly as expressed through their capacity to articulate, refine, and give voice to their own thinking. Many Indigenous communities have developed communication protocols that foreground authentic, relational, and reflective forms of expression. Practices such as Native American talking circles \cite{Kimmerer2015Braiding}, the Nahua concept of in xōchitl in cuīcatl (“flowery word”) \cite{leon1983cuicatl}, and yarning circles of Australian Aboriginal Peoples \cite{fricker2025yarning} emphasise speaking one’s truth in one’s own way, treating expression as inseparable from thinking, knowledge sharing, and the collective generation of understanding. 

GenAI -- particularly large language models -- offers significant opportunities for supporting communication and language revitalisation, including in Indigenous contexts \cite{Perera2025}. However, this promise also introduces a tension. When learners increasingly rely on AI systems to generate text, summarise ideas, or articulate arguments on their behalf, the cognitive work of formulating, refining, and expressing thought may be partially displaced. Emerging evidence that overreliance on GenAI can diminish critical thinking and reflective engagement \cite{Lee2025cognitive} raises concerns that sustained delegation of expressive labour to AI could weaken learners’ capacity to articulate ideas, reason through uncertainty, and develop voice as an expression of thought—capacities central both to Indigenous epistemologies and to learning more broadly.

\subsection{Unequal Access and Governance}
While GenAI holds potential to support education, access remains uneven. Worrell and Carlson \cite{worrell2025indigenous} note that many Indigenous communities continue to face structural barriers to potentially benefiting from digital technologies and AI, including limited internet connectivity, lack of access to digital infrastructure, and the remoteness of many communities. These inequities are compounded by concerns around data sovereignty, as most large-scale AI systems rely on externally controlled data, models, and governance structures that sit outside Indigenous authority. One promising response lies in the development of\textit{place-based AI agents}: locally deployed systems that operate on community-controlled infrastructure, are trained or constrained by locally approved data, and embed cultural protocols for access to knowledge \cite{Hsu2022Empowering}. Yet, these solutions must be approached cautiously. Worrell and Carlson \cite{worrell2025indigenous} also warn that systems that localise interfaces without transferring control over data, infrastructure, and governance risk reproducing colonial dynamics. Even well-intentioned AI tools can become extractive if Indigenous knowledge is encoded outside community authority or consent. Addressing unequal access therefore requires Indigenous-led governance, data sovereignty, and locally grounded protocols for managing access to sensitive knowledge, rather than the mere deployment of culturally branded technologies.

\section{Opportunities: Reciprocity and Design thinking as Design Resources}
The tensions outlined above are not arguments against AI in education, but invitations to rethink and redesign its purposes, governance, and use. The following opportunities respond directly to the tensions outlined above by re-framing AI design through reciprocity, participatory design, and Indigenous worldviews.

\subsection{Regenerative Innovation}
\textit{Regenerative innovation }offers an opportunity to address some of the challenges of AI in education by shifting design priorities from short-term optimisation toward long-term socio-ecological wellbeing. It has been defined as an approach to innovation that produces enduring, \textit{nature}-positive outcomes by strengthening the resilience of interconnected social and ecological systems, with value going first to the natural world and subsequently to communities and organisations \cite{Yadav2024RegenerativeInnovation}. This orientation directly responds to the environmental and extractive tensions discussed earlier. For example, an AIED system aligned with regenerative principles might operate on local or edge infrastructure \cite{sharma2025}, rely on community-curated data, and incorporate constraints on frequency or conditions of use to reduce unnecessary resource consumption while supporting meaningful learning.

From an Indigenous-aligned perspective, regenerative innovation extends conventional design criteria by introducing explicit obligations to relationships with people, place, and future generations, reframing innovation around whether technologies repair relationships, care for Country, and sustain collective wellbeing \cite{Nelson2023LinkedIn}. This reframes innovation not as a finished artefact but as an ongoing responsibility, requiring continued stewardship, reflection, and adaptation within the living systems it shapes. When applied to AI in education, regenerative innovation foregrounds reciprocity and care as core design resources, opening pathways for AI to support learning in ways that also sustain the communities and environments upon which learning depends. In turn, if AI systems are understood as infrastructures embedded in socio-ecological systems, then their governance must also reflect those relational dynamics.

One way to operationalise generative innovation we can look to “Ecological Corporations” that mimic biological systems \cite{turnbull2025sustaining}. Unlike hierarchical, profit-driven EdTech companies, these entities use “polycentric governance”, defined as distributed decision-making structures modelled on natural systems that function without a “Chief Executive Neuron.” Turnbull \cite{turnbull2025sustaining} argues that organisations designed in harmony with nature have “\textit{limited life}” and generate new “offspring” organisations, which prevents the concentration of power that leads to extractive monopolies. We propose applying this polycentric model to educational AI, treating these systems as common pool resources governed by education providers, educators and learners who use them, rather than as proprietary assets controlled by distant shareholders.

\subsection{AI as On-Demand Augmentation
}
A second opportunity is making deliberate decisions about when to use AI, and when not to. AI’s role in educational practice should be determined by pedagogical intent and situational context, rather than by default or continuous availability. Building on notions of intelligence augmentation \cite{Engelbart1962Augmenting}, educational AI should extend learners’ and educators’ cognitive capacities only when such support meaningfully enhances sense-making, reflection, or decision-making, rather than replacing productive struggle or human judgment. From this perspective, AI support should remain latent and be activated only when pedagogically justified. Decisions about \textit{when}, \textit{how}, and \textit{for whom} AI intervenes should therefore be treated as integral to task design. While these concerns echo longstanding work in AIED on scaffolding, learner agency, and collaborative learning \cite{Sharples03072023,LajoieLi2023TheoryDrivenAIED,MartinezMaldonado2023CollaborativeAIED} Importantly, treating AI as an on-demand resource also foregrounds reciprocity: each act of AI use draws on human labour, data, energy, and material resources, and therefore warrants deliberation rather than habitual or extractive deployment. Such an approach can reinforce human agency, pedagogical intent, and responsible engagement over technological convenience.

When AI is activated in alignment with clear pedagogical purposes, its effects can differ substantially from those observed under default or individualised use. For example, Sichterman et al. \cite{Sichterman2025} report that AI tools designed to visualise group performance and prompt shared reflection enhanced students’ socially shared regulation behaviours and fostered more effective co-regulation patterns than those found in groups without AI support. In this case, AI did not replace collaboration; rather, it was intentionally deployed to scaffold collective sense-making.

By contrast, when AI is continuously available and oriented toward individual optimisation, it risks reinforcing patterns of metacognitive offloading and learner isolation discussed earlier. Without deliberate boundaries and design intent, AI may displace productive struggle and reflective engagement rather than augment them. The opportunity, therefore, lies in positioning AI within relational learning ecologies in ways that intentionally deepen reflection, support collective regulation, and strengthen human agency.

\subsection{AI Literacy as Relational Literacy}

A third opportunity is in re-scoping the concept of AI literacy. 
Although AI literacy frameworks have expanded rapidly, most continue to prioritise technical understanding, effective use, and individual competency (see meta review: \cite{Zhang2025revies}). Recent competency-based approaches, such as UNESCO's AI competency framework for teachers, have begun to acknowledge ethical and environmental concerns, including environmental sustainability, yet these are typically framed as contextual knowledge rather than as design commitments that can shape when and whether AI should be used \cite{UNESCO2023AICompetency}. Critical AI and data literacy literature goes further by exposing how AI systems rely on power, data extraction, and human labour  \cite{Velander2024}. However, these approaches rarely articulate reciprocity as a guiding principle for AI literacy. Indigenous-led protocols and governance frameworks can offer a more fundamental reorientation, framing data, knowledge, and technology as relational rather than extractable, and insisting that AI use be justified through consent, accountability, and benefit to communities and Country \cite{Lewis2020IPAI}. This contrast reveals a persistent gap in AI literacy discourse: while environmental costs, data extractivism, and human labour are increasingly acknowledged, they remain peripheral to definitions of what it means to be AI literate, rather than central to how AI use is taught, designed, and evaluated.

The focus need not rest only on whether students are “ready” to use AI, but on whether AI systems, and the institutions adopting them, are ready for Indigenous students and knowledges \cite{Anderson02102023}. We might ask whether AI systems are “\textit{culturally ready}” for Indigenous learners. This reverses the deficit framing often applied to Indigenous learners and relocates responsibility within institutions and systems. This shifts the emphasis toward what is known about the readiness of schools and systems to meet Indigenous aspirations and invites critical reflection on whether those same institutions are culturally ready to integrate AI in ways that respect Indigenous knowledge. Importantly, designing AI systems that are responsive to Indigenous students should not be understood as a narrow or special-interest concern. Research in inclusive and universal design demonstrates that systems built with the needs of marginalised communities in mind often generate broader benefits for all users—the so-called “curb-cut effect” \cite{blackwell2017curb}, where design features created for accessibility enhance usability more generally. In this sense, ensuring that AI is culturally ready for Indigenous learners may strengthen its relational, pedagogical, and ethical robustness for all students.

\subsection{Indigenous AI Design and Participatory Design}
Meaningfully rethinking AI in education requires alternative ways of knowing and doing. Addressing the relational, material, and ethical challenges outlined above calls for engaging Indigenous knowledge systems that have sustained reciprocal relationships with land, knowledge, and community over millennia. Doing so, however, cannot be reduced to the selective adoption of concepts or frameworks. At least two steps are critical. First, there is a need to build stronger relationships grounded in respect, trust, and accountability among researchers, designers, educators, and the custodians of Indigenous knowledge. Second, these relationships must be accompanied by concrete action through the implementation of Indigenous-led principles and protocols for AI \cite{Lewis2020IPAI}. One practical starting point lies in participatory design and co-design approaches, such as design thinking, when these are undertaken in alignment with Indigenous values, governance, and practices. For example, the Kara Framework \cite{Jones2025} and the Both Ways Thinking Framework \cite{Nelson2023LinkedIn} offer visual and practice-oriented Indigenous design approaches grounded in Aboriginal and Torres Strait Islander ways of knowing, being, and doing, with the latter also drawing on Native American knowledge traditions and embedding kinship, accountability, and responsibility within culturally responsive design processes.

These commitments extend beyond governance into classroom practice, where Indigenous pedagogical frameworks offer models for structuring AI-enabled learning. One such model is Yunkaporta and Davies’ \cite{YunkaportaDavis2026} concept of Education as Embassy. Unlike standard cooperative learning, which can still retain competitive dynamics, their “\textit{Ko-Learning}” approach frames the classroom as an embassy space, where power does not sit in individuals, institutions, or coercive behaviours, but in a foundation of good relation. In a Ko-Learning AI classroom, each learner is understood as a node in a complex system, mirroring the relational dynamics of land and community. The pedagogy structures interaction through specific kinship relations: -I- (autonomy), -us-two- (negotiated pairs), and -us-all- (whole team embassy). Applying these protocols to AI design, participation, and literacy shifts practice away from individualised consumption of technology towards symbiotic relations, where transactions are understood as give-and-take within a balanced network of flows. This prevents AI from becoming an extractive force and instead situates it as part of a reciprocal knowledge ecosystem.

Moreover, in designing AI systems that respect Indigenous worldviews, we can adopt Yunkaporta and Shillingsworth’s \cite{YunkaportaShillingsworth2020} framework: Respect (Axiology) → Connect (Ontology) → Reflect (Epistemology) → Direct (Methodology), with “Spirit and Integrity – Respect” as the central core value that binds all others. This framework extends beyond Western notions of research integrity. Here, “Spirit” speaks to continuity between past, present, and future generations, while “Integrity” refers to the respectful behaviours that hold these values together. For AI in education, this means every algorithmic decision should be assessed not only for accuracy or efficiency, but for its alignment with the continuity of cultural knowledge and with our intergenerational obligations to future learners.

\section{Concluding Remarks: Future Pedagogical Directions for the Next Seven Generations}
Across various Indigenous worldviews, there is a shared ethical responsibility to become a\textit{ good ancestor}, that is, to make decisions in the present that sustain life, knowledge, and relationships for those who will come after. In some Native American traditions, this is made explicit through the practice of considering the consequences of actions for the next seven generations \cite{Munroe04052022}, a principle that is particularly salient when designing systems, infrastructures, and socio-technical arrangements with long-term effects \cite{Lewis2020IPAI}. 

Bringing this cyclical and relational orientation into AI design offers a concrete foundation for technologies that prioritise the sustained wellbeing of communities and environments, rather than short-term performance or extraction. This orientation is not misaligned from the aims of many researchers and designers who seek to deploy AI for educational and social benefit. Yet, the economic imperatives, concentration of power, and extractive business models that currently drive much of the AI ecosystem (largely controlled by Big Tech) systematically constrain designs that centre reciprocity, care, and collective responsibility, raising unresolved questions of sovereignty over data, knowledge, and technological futures \cite{Iazzolino01102024}. 

This position paper has therefore foregrounded challenges rather than prescriptive solutions, reflecting both the pace of AI-driven change and the limits of addressing these challenges through the same modes of thinking that have contributed to social and ecological crises. Learning from and standing in solidarity with Indigenous peoples, whose knowledge systems  and governing have sustained reciprocal relationships with land and community over millennia, is thus not only a matter of justice, but a necessary condition for reimagining education itself. If education is, as long argued, a social, relational, and constructive practice through which knowledge is co-created in responsibility to others \cite{biesta2015beyond,engestrom1987learning,freire1970pedagogy}, then rethinking AI in education must begin not with optimisation or efficiency, but with the relations we choose to sustain -- with self, with others, with kin, and with Country -- and the futures we are willing to be accountable for.
%
%
\bibliographystyle{splncs04}
\bibliography{mybibliography}
\end{document}